\documentclass[twocolumn,preprintnumbers,amsmath,amssymb,superscriptaddress,nofootinbib,english]{revtex4-1}
\pdfoutput=1
\usepackage{times,amsmath,amsfonts,amssymb,epstopdf}
\usepackage{graphicx}
\usepackage{dcolumn}
\usepackage{bm}
\usepackage{enumerate}
\usepackage{epsfig}
\usepackage{graphicx}
\usepackage{hyperref}
\usepackage[usenames]{color}
\usepackage{url}
\usepackage[normalem]{ulem}
\usepackage[T1]{fontenc}
\usepackage{amsmath}

\usepackage[usenames]{color}

\def\sc{\scriptscriptstyle}
\def\be{\begin{equation}}
\def\ee{\end{equation}}
\def\ben{\begin{eqnarray}}
\def\een{\end{eqnarray}}
\def\ba{\begin{array}}
\def\ea{\end{array}}

\newcommand{\bq}{\begin{eqnarray}}
\newcommand{\eq}{\end{eqnarray}}
\newcommand{\bes}{\begin{subequations}}
\newcommand{\ees}{\end{subequations}}

\begin{document}
\newcommand{\half}{{\textstyle\frac{1}{2}}}
\allowdisplaybreaks[3]
\def\triangledown{\nabla}
\def\grad3{\hat{\nabla}}
\def\a{\alpha}
\def\b{\beta}
\def\g{\gamma}\def\G{\Gamma}
\def\d{\delta}\def\D{\Delta}
\def\ep{\epsilon}
\def\et{\eta}
\def\z{\zeta}
\def\t{\theta}\def\T{\Theta}
\def\l{\lambda}\def\L{\Lambda}
\def\m{\mu}
\def\f{\phi}\def\F{\Phi}
\def\n{\nu}
\def\r{\rho}
\def\s{\sigma}\def\S{\Sigma}
\def\ta{\tau}
\def\x{\chi}
\def\o{\omega}\def\O{\Omega}
\def\k{\kappa}
\def\pa {\partial}
\def\ov{\over}
\def\br{\\}
\def\ud{\underline}

\def\lcdm{\Lambda{\rm CDM}}
\def\qcdm{{\rm QCDM}}
\def\nloc{R\Box^{-2}R}
\def\msun{M_{\odot}/h}
\def\dw{f(X)}
\def\costhe{{\rm cos}\theta}
\def\sinthe{{\rm sin}\theta}
\def\cosphi{{\rm cos}\varphi}
\def\sinphi{{\rm sin}\varphi}
\def\sintwothe{{\rm sin}^2\theta}
\def\costwothe{{\rm cos}^2\theta}
\def\sintwophi{{\rm sin}^2\varphi}
\def\costwophi{{\rm cos}^2\varphi}
\def\boxa{L = 250\ {\rm Mpc}/h}
\def\boxb{L = 100\ {\rm Mpc}/h}
\def\boxc{L = 400\ {\rm Mpc}/h}
\def\ttwo{h_c \leq 0.03\ \left[{\rm Mpc}/h\right]}
\def\tfour{h_c \leq 0.12\ \left[{\rm Mpc}/h\right]}
\def\tfive{h_c \leq 0.24\ \left[{\rm Mpc}/h\right]}

\def\pxpx{\left[\partial_x\partial_x\Phi\right]}
\def\pypy{\left[\partial_y\partial_y\Phi\right]}
\def\pzpz{\left[\partial_z\partial_z\Phi\right]}
\def\pxpy{\left[\partial_x\partial_y\Phi\right]}
\def\pxpz{\left[\partial_x\partial_z\Phi\right]}
\def\pypz{\left[\partial_y\partial_z\Phi\right]}

\newcommand{\ramses}{{\sc ramses}}
\newcommand{\isis}{{\sc{isis}}}
\newcommand{\isisns}{{\sc{isis-nonstatic}}}
\newcommand{\ecosmog}{{\sc ecosmog}}
\newcommand{\dgpm}{{\sc dgpm}}

\title{Speeding up N-body simulations of modified gravity: Vainshtein screening models}

\author{Alexandre Barreira}
\email[Electronic address: ]{barreira@mpa-garching.mpg.de}
\affiliation{Institute for Computational Cosmology, Department of Physics, Durham University, Durham DH1 3LE, U.K.}
\affiliation{Institute for Particle Physics Phenomenology, Department of Physics, Durham University, Durham DH1 3LE, U.K.}
\affiliation{Max-Planck-Institut f{\"u}r Astrophysik, Karl-Schwarzschild-Str. 1, 85748 Garching, Germany}

\author{Sownak Bose}
\affiliation{Institute for Computational Cosmology, Department of Physics, Durham University, Durham DH1 3LE, U.K.}

\author{Baojiu Li}
\affiliation{Institute for Computational Cosmology, Department of Physics, Durham University, Durham DH1 3LE, U.K.}

\begin{abstract}

We introduce and demonstrate the power of a method to speed up current iterative techniques for N-body modified gravity simulations. Our method is based on the observation that the accuracy of the final result is not compromised if the calculation of the fifth force becomes less accurate, but substantially faster, in high-density regions where it is weak due to screening. We focus on the nDGP model which employs Vainshtein screening, and test our method by running AMR simulations in which the solutions on the finer levels of the mesh (high density) are not obtained iteratively, but instead interpolated from coarser levels. We show that the impact this has on the matter power spectrum is below $1\%$ for $k < 5h/{\rm Mpc}$ at $z = 0$, and even smaller at higher redshift. The impact on halo properties is also small ($\lesssim 3\%$ for abundance, profiles, mass; and $\lesssim 0.05\%$ for positions and velocities). The method can boost the performance of modified gravity simulations by more than a factor of 10, which allows them to be pushed to resolution levels that were previously hard to achieve.

\end{abstract}

\maketitle
\section{Introduction}\label{sec:intro}

One of the main goals of current (BOSS \cite{2013MNRAS.429.1514S}, CFHTLenS \cite{2012MNRAS.427..146H, 2013MNRAS.429.2249S}, DES \cite{Abbott:2015swa}) and upcoming (DESI \cite{2013arXiv1308.0847L}, LSST \cite{2012arXiv1211.0310L}, Euclid \cite{2011arXiv1110.3193L}) observational missions is to test the gravitational law on cosmological scales. In some of the most popular alternative gravity models (see e.g.~Refs.~\cite{2012PhR...513....1C, Joyce:2014kja} for reviews), the departures from General Relativity (GR) are typically caused by a scalar field that mediates an additional {\it fifth force} felt by the matter fields. These models must ensure that the amplitude of the fifth force is sufficiently small in regimes where gravity is very well tested, like in the Solar System \cite{Will:2014xja}. On the other hand, these models should also leave space for sizeable fifth forces in regimes where gravity is not well tested yet, like in cosmology or other astrophysical environments, if they are to be distinguishable from GR \cite{Jain:2007yk, Jain:2013wgs, 2015arXiv150404623K, Sakstein:2014nfa, Sakstein:2015zoa}. These two conditions are typically reconciled with screening mechanisms \cite{Brax:2013ida}, which are dynamical effects that arise from nonlinear terms in the equations and that naturally suppress the size of the modifications to gravity in regions where the density is high relative to the cosmological average, like in the Solar System \footnote{In the literature (and as we do here), it is common to describe screening mechanisms as being triggered by higher values of the matter density, but this constitutes an abuse of language. For instance, in the Solar System, the density between the Sun and the planets is very low, but one still requires the screening to operate. A more accurate formulation is that the screening efficiency is higher in regions of high potential, acceleration or curvature.}. The idea of modifying gravity on large scales can also help to explain the observed accelerated expansion of the Universe, which provides extra motivation for these studies.

Owing to the screening mechanism, the relative size of the fifth force to normal gravity becomes dependent on the detailed distribution of matter on small (nonlinear) scales. The necessity to model the nonlinear clustering of matter accurately means that one has to recourse to N-body simulations. The current state-of-the-art techniques \cite{2012JCAP...01..051L, baojiudgp, 2013MNRAS.436..348P, 2014AA...562A..78L} are based on Adaptive Mesh Refinement (AMR) methods, which make use of a mesh that covers the simulation volume and whose cells are made finer in regions of high density. This ensures high enough resolution in high-density regions, whilst limiting computational costs in low-density regions where the resolution can be lower. The nonlinear equation of the scalar field is discretised onto the AMR mesh and solved on every refinement level using iterative algorithms. The development of the first parallelizable AMR code of modified gravity was achieved with the {\tt ECOSMOG} code \cite{2012JCAP...01..051L, baojiudgp}, which is based on the publicly-available {\tt RAMSES} code \cite{romain}. Subsequent efforts resulted in the development of the {\tt MG-GADGET} \cite{2013MNRAS.436..348P} and {\tt ISIS} \cite{2014AA...562A..78L} codes. The results from these three codes (and another non-AMR code \cite{2009PhRvD..80d3001S, 2009PhRvD..80l3003S}) were recently compared in Ref.~\cite{codecomp}, where it was found that the codes agree extremely well in their predictions for the relative deviation from standard $\lcdm$. For instance, the code differences in the power spectrum are kept below $1\%$ for $k \lesssim 5h/ {\rm Mpc}$ at $z = 0$.

It is unquestionable that N-body simulations of modified gravity have played and continue to play a significant role in our understanding of the types of observational signatures that can be present on nonlinear scales. However, there is one unnapealing aspect about current N-body methods for modified gravity, which is that, although remarkably efficient given the complexity of the problem, they are still significantly more time consuming than $\lcdm$ simulations. To get a rough feeling of the slowdown, for simulation boxes with side $L = 250 - 500\ {\rm Mpc}/h$ and $N_p = 256^3 - 512^3$ particles, the modified gravity simulations can typically take $5-15$ times longer than $\lcdm$. Moreover, the slowdown tends to become more significant with increased particle resolution. This limitation has made it difficult to run simulations of modified gravity with the resolution and size that is currently possible with $\lcdm$. The existence of such simulations can be important in the calibration of observational studies (e.g.~via generation of covariance matrices or emulators \cite{2014ApJ...780..111H}), or in further theoretical studies such as hydrodynamical simulations (see Refs.~\cite{2014MNRAS.440..833A, 2015MNRAS.448.2275A, 2015MNRAS.449.3635H, 2015arXiv150506803H, 2015arXiv150807350H} for some firsts steps in this direction) or even galaxy formation simulations in alternative gravity cosmologies.

There has been some effort to obtain fast predictions of modified gravity on small scales. One of these are predictions based on the semi-analytical spherical collapse and halo model \cite{2012MNRAS.421.1431L, 2013JCAP...11..056B, 2013PhRvD..88h4015K,2014JCAP...03..021L, 2014JCAP...04..029B}. These calculations are very fast and able to provide a qualitative feeling for the types of effects. However, they typically require some level of calibration against N-body simulation results \cite{2014JCAP...03..021L, 2014JCAP...04..029B}, and normally fail to meet the accuracy requirements of current and future surveys. Another interesting approach is that of Ref.~\cite{2015PhRvD..91l3507W}, who proposed a numerical method to simulate modified gravity without significant numerical overheads relative to $\lcdm$. In their method, instead of solving the nonlinear equation of the scalar field, the code solves a linearized version weighted by a screening factor that can be calculated analytically assuming spherical symmetry (see also Ref.~\cite{2009PhRvD..80f4023K} for a similar approach). Since the scalar field equation becomes linear, it can then be solved using the same fast methods that are used to solve the linear Poisson equation in standard gravity. In Ref.~\cite{2015PhRvD..91l3507W}, the authors applied this technique to different models and found that the fast simulations can recover the power spectrum of the full simulation to within $\sim 3\%$ for $k \lesssim 1 h/{\rm Mpc}$ (these figures vary slightly from model to model). More recently, Ref.~\cite{2015MNRAS.452.4203M} proposed a method in which the output of a standard gravity simulation is rescaled in order to reproduce the result from a different cosmology \cite{2010MNRAS.405..143A}, which can be a modified gravity one. In Ref.~\cite{2015MNRAS.452.4203M}, is was shown that the method can reproduce the power spectrum of the Hu-Sawicki $f(R)$ model \cite{Hu:2007nk} with $\sim 3\%$ accuracy, up to $k \sim 0.1h/{\rm Mpc}$ or $k \sim 1h/{\rm Mpc}$ (depending on the level of modelling involved, see Ref.~\cite{2015MNRAS.452.4203M} for the details).

In this paper, we propose another method to speed up current N-body simulations of modified gravity, but whose accuracy extends further into the nonlinear regime compared to previous efforts. Our method is based on the observation that it may be unnecessary to spend a lot of computational time solving for the scalar field in highly screened regions, where the fifth force is very weak. This is in the sense that it is possible to afford an inaccurate fifth force solution where the fifth force contribution to the total force is negligible. We consider simulations of the Dvali-Gabadadze-Porrati \cite{Dvali:2000hr} (DGP) gravity model to illustrate the performance of our method, which we implement by truncating the iterations of the scalar field on high refinement levels of the AMR mesh, where the fifth force is weaker due to screening. One of our main results is that our method can result in an improvement of a factor of $\sim 10$ in the performance of the simulations, with barely any loss in accuracy, at least down to relatively small scales $k \approx 5h/{\rm Mpc}$.

The outline of this paper is as follows. In Sec.~\ref{sec:dgp}, we introduce the DGP gravity model and the Vainshtein screening mechanism \cite{Vainshtein1972393, Babichev:2013usa, Koyama:2013paa, 2015PhRvD..91l4066K} that operates in it. In Sec.~\ref{sec:method}, we summarise the main features of state-of-the-art iterative methods for modified gravity simulations and explain how our method can be used to speed them up. We also discuss how the implementation of the speed-up method depends on the properties of the screening mechanism at play. Section \ref{sec:results} is devoted to demonstrating the validity of our method by comparing the results from the truncated and full simulations. We compare the simulation results in their predictions for the dependence of the size fifth force on the AMR refinement level in Sec.~\ref{sec:ratio}; the matter power spectrum in Sec.~\ref{sec:timeevo} and halo counts, concentration, spin and profiles in Sec.~\ref{sec:counts}.  In Sec.~\ref{sec:props}, compare a number of properties of haloes in the truncated runs to matched haloes in the full simulations. Finally, we summarize and conclude in Sec.~\ref{sec:conc}.

Unless otherwise specified, we work with units where $c = 1$, where $c$ is the speed of light.

\section{The DGP gravity model}\label{sec:dgp}

The braneworld DGP model \cite{Dvali:2000hr} is amongst the most thoroughly studied modified gravity models in the nonlinear regime with N-body simulations \cite{2009PhRvD..80d3001S, 2009PhRvD..80l3003S, baojiudgp, 2012PhRvL.109e1301L, 2013MNRAS.436...89R, 2013PhRvD..88h4029W, 2014MNRAS.445.1885Z, 2014JCAP...07..058F, 2015JCAP...07..049F} and it is perhaps the leading representative of models that employ the Vainshtein screening mechanism. In this model, standard GR and the known matter fields are defined on a four-dimensional brane that is embedded in a five-dimensional bulk spacetime containing a five-dimensional generalization of GR. The action of the model can be written as

\bq\label{eq:dgpaction}
S &=& \int_{\rm brane}\!\!\! {\rm d}^4x \sqrt{-g} \left(\frac{R}{16\pi G} \right) 
+ \int {\rm d}^5x \sqrt{-g^{(5)}} \left(\frac{R^{(5)}}{16\pi G^{(5)}}\right) \nonumber\\
&& +S_m(g_{\mu \nu}, \psi_i),
\eq
where $g^{(5)}$ and $g$ are the determinants of the metric of the bulk ($g_{\mu\nu}^{(5)}$) and brane ($g_{\mu\nu}$), respectively, and $R^{(5)}$ and $R$ are their Ricci scalars. The matter fields are denoted by $\psi_i$ and their action $S_m$ belongs to the four-dimensional part of the model. The ratio of the two gravitational strengths, $G^{(5)}$ and $G$, is a parameter of the model known as the crossover scale, $r_c$,
\bq\label{eq:rc}
r_c = \frac{1}{2}\frac{G^{(5)}}{G}.
\eq

The expansion rate in this model can be writen as (e.g.~\cite{2007PhRvD..75h4040K, 2006JCAP...01..016K})
\bq\label{eq:dgpH}
H(a) = H_0\sqrt{\Omega_{m0}a^{-3} + \Omega_{rc}} \pm \sqrt{\Omega_{rc}},
\eq
where $a$ is the cosmological scale factor, $H_0 = 100h\ {\rm km/s/Mpc}$ is the present-day Hubble expansion rate, $\Omega_{m0} = \bar{\rho}_{m0}8\pi G/(3H_0^2)$ is the present-day fractional matter density, $\bar{\rho}_{m0}$ is the present-day background value of the matter density, $\rho_{m}$, and $\Omega_{rc} = 1/(4H_0^2r_c^2)$ (the subscripts $_0$ denote present-day values). This model has two branches of solutions, characterized by the sign of the second term on the right-hand side in Eq.~(\ref{eq:dgpH}). The so-called {\it self-accelerating branch} (sDGP) which corresponds to the $(+)$ sign, is particularly appealing as it allows for solutions in which the acceleration of the Universe arises without adding any explicit dark energy component such as a cosmological constant or a smooth scalar field. However, this branch is known to be plaged by ghost problems (unstable degrees of freedom without a well defined minimum energy state) \citep{2003JHEP...09..029L,2004JHEP...06..059N, 2007CQGra..24R.231K}. In addition to these theoretical instabilities, the self-accelerating branch is in severe tension with CMB and supernovae data \citep{2008PhRvD..78j3509F}. For these reasons, most of the cosmological studies of DGP gravity have focused on the so-called {\it normal branch}, nDGP, which is characterized by the $(-)$ sign. This branch requires the addition of an explicit dark energy component, $\rho_{de}$, on the brane. This gives rise to a term $\rho_{de}(a)$ inside the square root in the first term on the right-hand side of Eq.~(\ref{eq:dgpH}), which can be tuned so that the expansion rate in the nDGP model matches exactly that of $\Lambda$CDM \cite{2009PhRvD..80l3003S}. This match is not mandatory, nor strictly required by observations, but it does help in comparing the simulation results to $\Lambda$CDM. This is in the sense that the changes in the observables are then governed by the extra force (see below), and not by changes to the background expansion rate. Naturally, the necessity of a dark energy component removes some of the merit of the theory, but yields a self-consistent model that can (and has) prove very useful in learning about modified gravity signatures on cosmological scales. From hereon, we focus on the normal branch. 

On scales much smaller than the horizon ($\ll c/H$) and the crossover scale ($\ll r_c$), the formation of structure in the nDGP model is governed by the modified Poisson equation,
\bq\label{eq:modpoisson}
\nabla^2\Psi = 4\pi G a^2 \delta\rho_m + \frac{1}{2}\nabla^2\varphi,
\eq
where $\delta\rho_m = \rho_m - \bar{\rho}_{m}$ is the matter density perturbation and $\varphi$ is a scalar degree of freedom associated with the bending modes of the brane and whose equation of motion is given by
\bq\label{eq:eomvarphi}
\nabla^2\varphi + \frac{r_c^2}{3\beta(a) a^2} \left[\left(\nabla^2\varphi\right)^2 - \left(\nabla_i\nabla_j\varphi\right)^2\right] = \frac{8\pi G}{3\beta(a)}a^2 \delta\rho_m. \nonumber \\
\eq
The above equations correspond to a perturbed Friedmann-Robertson-Walker (FRW) metric on the brane
\bq\label{eq:metric}
{\rm d}s^2 = \left(1 + 2\Psi\right){\rm d}t^2 - a(t)^2\left(1 - 2\Phi\right)\gamma_{ij}{\rm d}x^i{\rm d}x^j,
\eq
and are obtained by employing the weak-field limit, $\Psi, \Phi, \varphi \ll 1$ (assuming the same boundary conditions for the gravitational potentials and the scalar field) and the quasi-static approximation, which amounts to neglecting time-derivatives of $\varphi$ over spatial ones. The validity of this approximation has been verified in Refs.~\cite{2009PhRvD..80d3001S, 2014PhRvD..90l4035B, 2015arXiv150503539W}. In Eqs.~(\ref{eq:modpoisson}) and (\ref{eq:eomvarphi}), $\beta(a)$ is a function of time given by
\bq\label{eq:beta}
\beta(a) = 1 + 2H(a)r_c\left(1 + \frac{\dot{H}(a)}{3H(a)^2}\right),
\eq
where the dot denotes a derivative w.r.t.~physical time $t$. In DGP gravity, $\varphi$ affects dynamics of massive particles alone, $\Psi = \Psi_{\rm GR} + \varphi$, without contributing directly to the lensing potential, $\Phi_{\rm len} = \left(\Phi + \Psi\right)/2 = \Phi_{\rm GR}$.

\subsection{The Vainshtein screening mechanism}

\begin{figure}
	\centering
	\includegraphics[scale=0.50]{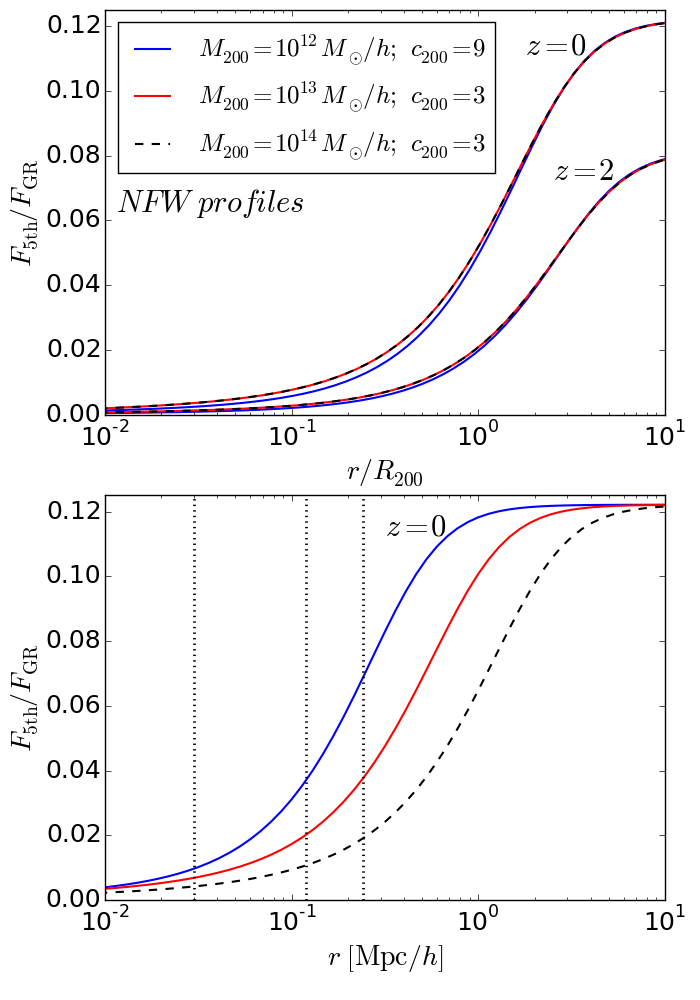}
	\caption{Fifth force to normal gravity ratio for NFW halo profiles in the nDGP model at $z = 0$ and $z=2$ and for different values of $M_{200}$ and $c_{200}$, as labelled. The model parameters are the same as the simulations used in this paper (cf.~Sec.~\ref{sec:sumsim}). In the upper panel, this is plotted against $r/R_{200}$. The lower panel shows only the $z=0$ result and the $x$-axis is $r$ to highlight the different halo sizes for different halo masses. The three vertical lines in the lower panel indicate the grid sizes of different truncation criteria of the scalar field iterations in the N-body code (cf.~Secs.~\ref{sec:speed} and \ref{sec:sumsim}).}
\label{fig:nfw}
\end{figure}

To illustrate the Vainshtein screening effect, it is best to assume spherical symmetry. This way, integrating Eq.~(\ref{eq:eomvarphi}) as $4\pi\int r^2{\rm d}r$, we get
\bq\label{eq:eomvarphi_sph}
\frac{\varphi,_r}{r} + \frac{2r_c^2}{3\beta}\left(\frac{\varphi,_r}{r}\right)^2 = \frac{2}{3\beta}\frac{GM(<r)}{r^3},
\eq
where the comma denotes partial differentiation w.r.t.~the comoving radial coordinate, $r$, and $M(<r) = 4\pi\int_0^r \xi^2\rho(\xi){\rm d}\xi$ is the mass enclosed by radius $r$. This equation can be solved analytically to yield
\bq\label{eq:eomvarphi_sph2}
\varphi,_r = \frac{4}{3\beta}\left(\frac{r}{r_V}\right)^3\left[-1 + \sqrt{1 + \left(\frac{r_V}{r}\right)^3}\right]\frac{GM(r)}{r^2},
\eq
in which 
\bq\label{eq:rv}
r_V(r) = \left(\frac{16r_c^2GM(r)}{9\beta^2}\right)^{1/3}
\eq
is a distance scale known as the {\it Vainshtein radius}. At a given radius $r$, $r_V(r)$ sets the scale below which the size of the fifth force, $F_{5 \rm th} = \varphi,_r/2$ starts to become suppressed. For concreteness, consider a top-hat with size $R_{\rm TH}$ and mass $M_{\rm TH}$. For $r \gg r_V(r) > R_{\rm TH}$,
\bq\label{eq:forcelarge}
\frac{\varphi,_r}{2} \approx \frac{1}{3\beta}\frac{G M(<r)}{r^2} = \frac{1}{3\beta}F_{\rm GR}.
\eq
That is, the fifth force can be a sizeable fraction of the GR force, $F_{\rm GR}$. However, for  $R_{\rm TH} < r \ll r_V(r)$, we have
\bq\label{eq:forcesmall}
\frac{F_{5\rm th}}{F_{GR}} \rightarrow 0, \ \ \ \ \ \ {\rm as} \ \ \ \ \ \frac{r}{r_V} \rightarrow 0.
\eq
As another example, we show in Fig.~\ref{fig:nfw} the profile of the ratio $F_{5\rm th}/F_{\rm GR}$ for NFW haloes
\bq\label{eq:nfw}
\rho_{\rm NFW} &=& \frac{\rho_s}{\left(r/r_s\right)\left(1 + r/r_s\right)^2},
\eq
where the parameters $\rho_s$ and $r_s$ relate to the halo mass $M_{200}$ and halo concentration $c_{200}$ as
\bq\label{eq:nfwparams}
\rho_s &=& \frac{200}{3}\rho_cc_{200}^3\left[{\rm ln}(1+c_{200}) - \frac{c_{200}}{1 + c_{200}}\right]^{-1}, \nonumber \\
r_s &=& R_{200}/c_{200}.
\eq
The radial coordinate in the upper panel of Fig.~\ref{fig:nfw} is scaled by $R_{200}$,
\bq\label{eq:r200}
R_{200} &=& \left(\frac{3M_{200}}{4\pi 200 \rho_c(z)}\right)^{-3},
\eq
which is the radius above which the mean density of the halo drops below $200$ times the critical matter density, $\rho_c(z) = 3H(z)^2/(8\pi G)$, at a given redshift. The result is shown for three combinations of $M_{200}$ and $c_{200}$ at $z=0$ and $z=2$, as labelled. The figure shows that the suppresional effects of the screening mechanism start to become important on scales $\lesssim 10R_{200}$. Moreover, the figure also illustrates the known result that the fifth force to normal gravity ratio is independent of $M_{200}$ when plotted as a function of $r/R_{200}$ \footnote{This can be checked by noting that Eq.~(\ref{eq:eomvarphi_sph2}) depends on the combination $r_V^3/r^3$, which depends only on $r/R_{200}$ and $c_{200}$ (and not $M_{200}$) for NFW halos.}. This is why the dashed black and solid red curves are overlapping in the upper panel of Fig.~\ref{fig:nfw}. The concentration has a slight impact on the ratio $F_{5{\rm th}}/F_{\rm GR}$. Namely, the higher $c_{200}$, the higher $M(r)$, and hence $r_V$, which strengthens the screening mechanism (this is why the blue curves have a lower amplitude). We shall refer to this figure when discussing the numerical method that we propose in this paper.

\section{Faster simulations of modified gravity}\label{sec:method}

We start this section by outlining the commonly adopted algorithms that are used to perform N-body simulations of modified gravity. We then explain the main premises and implementation of our method.

\subsection{Iterative methods with multigrid acceleration}\label{sec:itermet}

In modified gravity simulations, the goal is to determine the total modified force, $-\nabla\Psi$, where the potential $\Psi$ normally satisfies a modified Poisson equation that can be written as
\bq\label{eq:poissongen}
\nabla^2\Psi = \nabla^2\Psi_{\rm GR} + f(\varphi, \nabla\varphi, \nabla^2\varphi),
\eq
where $f$ is some function of a scalar field and/or its derivatives. For the nDGP model, $f = \nabla^2\varphi/2$ (cf.~Eq.~(\ref{eq:modpoisson})). The scalar field is usually governed by a nonlinear Poisson-like equation, whose form can be cast generically as
\bq\label{eq:eomgen}
\mathcal{L}\left[\varphi\right] = \mathcal{S}\left(\delta\rho_m\right),
\eq
where $\mathcal{L}$ is some derivative operator acting on $\varphi$ and $\mathcal{S}$ is a function of the density perturbation, which sources the scalar field. Equation (\ref{eq:eomgen}) can be contrasted with the nDGP equation, Eq.~(\ref{eq:eomvarphi}). The N-body particles are evolved according to
\bq\label{eq:geodesic}
\ddot{x} + 2H\dot{x} = -\nabla\Psi,
\eq
which is as in standard GR simulations, just with a modified dynamical potential, $\Psi$.

The challenging and time consuming part of N-body simulations of modified gravity is to solve Eq.~(\ref{eq:eomgen}). The method employed by {\tt ECOSMOG} code (and adopted also by other state-of-the-art codes \cite{codecomp}) consists of discretising the equation on an AMR mesh and then, at each simulation time step, iterating over the values of the scalar field on all AMR cells until a certain convergence criterion is met. Explicitly, one solves the equation 
\bq\label{eq:eomgen-disc}
\mathcal{T}_{ijk}^l \equiv \mathcal{L}\left[\varphi\right]_{ijk}^{l} - S_{ijk}^{l} = 0,
\eq
where $l$ labels the refinement level of the AMR structure and $\left\{ijk\right\}$ labels each cell on refinement level $l$. The iterations can be performed using the {\it Newton-Raphson} method in which the new value of the scalar field in each iteration is given in terms of the current one as 
\bq\label{eq:newton}
\varphi_{ijk}^{new, l} = \varphi_{ijk}^{l} - \frac{\mathcal{T}_{ijk}^l}{\partial \mathcal{T}_{ijk}^l/\partial\varphi_{ijk}^l}.
\eq
In {\tt ECOSMOG}, the iterations are performed on an AMR level basis. The code first solves the equation on the domain (unrefined) grid which has periodic boundary conditions. The initial guess for the iterations can be set to zero or to the value obtained at the previous time step.  The boundary conditions on higher refinement levels are fixed and are obtained via interpolation from coarser levels. The initial guess for the iterations on the refinements is also obtained via interpolation from the coarser solutions. As in {\tt RAMSES}, in {\tt ECOSMOG} all the fields ($\delta\rho$, $\Phi$ and $\varphi$) are evaluated at the center of the AMR cells. The interested reader can find the discretised version of the operators $\mathcal{T}_{ijk}$ and $\partial T_{ijk}^l/\partial\varphi_{ijk}^l$ for the DGP model as Eqs.~(29) and (32), respectively, in Ref.~\cite{baojiudgp} (see also Refs.~\cite{2013JCAP...10..027B, 2013JCAP...11..012L} for Galileon model simulations in {\tt ECOSMOG}, which also employ Vainshtein screening).

In such an iteration scheme, it is well known that the error reduces quickly during the first few iterations, but less so afterwards. This is due to the fact that the Fourier modes of the solution with wavelength larger than the grid size have poor convergence properties. A way around this problem consists in making use of a hierarchy of coarser grids in a method known as {\it multigrid acceleration}. In this method, after the first few iterations on level $l$, the equation is interpolated onto a coarser grid, $l - 1$, where further iterations take place to suppress the larger wavelength modes of the error that could not be quickly suppressed in the finer grid. This {\it coarsification} can be done for several coarse levels, $l-2, l-3$ ... etc. The equation is then interpolated back onto the finer (original) level, $l$, and, if convergence is not yet reached, then further coarsifications take place. These processes of going {up and down} in resolution are called V-cycles. We refer the reader to Sec.~3.~2.~of Ref.~\cite{codecomp} for a summary of the multigrid acceleration method in the context of modified gravity simulations.

The multigrid method plays a crucial role in giving N-body codes of modified gravity an acceptable performance. However, the fact that the iterations of Eq.~(\ref{eq:newton}) need to be performed for (i) every simulation time step; (ii) all cells on a given AMR level; and (iii) for every AMR level, still makes N-body simulations of modified gravity considerably slower than its GR counter parts. For instance, for the box size $L=250\ {\rm Mpc}/h$ and particle number $N_p = 512^3$ used in Ref.~\cite{codecomp}, the simulations of modified gravity can still be up to $5-10$ times slower than GR. These figures become worse with increasing particle resolution, as seen, for example, in the higher resolution efforts made by Refs.~\cite{2015MNRAS.448.2275A, 2015MNRAS.452.3179S}. In the remainder of this section, we explain how modified gravity simulations can be sped up significantly by relaxing condition (iii) above, without compromising the accuracy of the result.

\subsection{The speed-up method: truncation of the iterations in highly refined/screened regions}\label{sec:speed}

\begin{figure}
	\centering
	\includegraphics[scale=0.30]{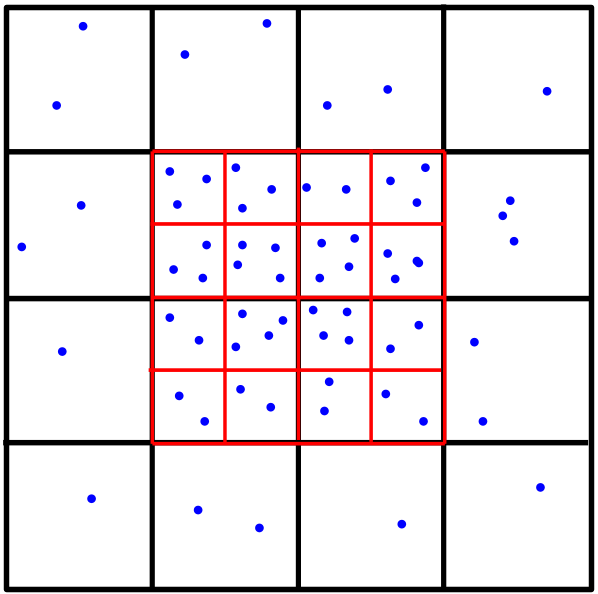}
	\caption{Two-dimensional illustration of a refined region in the AMR structure. The idea of the speed-up method is that, provided the screening mechanism is already at play, the fifth force at the center of the fine cells (red) does not need to be solved explicitly and expensively, but rather it can be interpolated from the values of the fifth force that have already been calculated at the center of the coarse cells (black). The blue dots display an illustrative particle distribution. Note that the fifth force solutions in the coarse level capture also the fact that the matter density is higher in the inner region of the domain shown.}
\label{fig:amr}
\end{figure}

The speed-up method that we propose here takes advantage of the fact that the amplitude of the fifth force in Vainshtein screening models decreases with increasing refinement level (where the density is higher) because of the increased efficiency of the screening mechanism. This means that by solving Eq.~(\ref{eq:newton}) on the refinements, {\it the code can spend a large amount of time to get an accurate result to a quantity that is only a very small correction to the total force}. In the method that we propose and test in this paper, instead of iteratively solving for the scalar field on all refinements, we truncate the iterations above a given refinement level where we take the solution to be the initial guess that is interpolated from the coarser levels. One of the main results of this paper is that {\it these truncations can significanly boost the performance of the modified gravity N-body code, with little sacrifice in accuracy}.

Figure \ref{fig:amr} shows a two-dimensional sketch of a refined region that helps to visualize the implementation of our method. The figure shows only one refinement level for simplicity, but the reasoning is straighforwardly generalizable to more refinement levels. To compute the fifth force, one first iterates the scalar field values on the coarser level (black) until convergence. The fifth force in the nDGP model, $F_{5\rm th} = \nabla\varphi/2$, is then given by finite-diferencing the scalar field values evaluated at the center of the coarse cells. It is important to note that the higher matter density/screening efficiency at the center of the domain shown is captured also by the coarse level, and that this translates into a suppresion of the fifth force there. Then, instead of performing the same iterations on the fine level (red), we can truncate them and simply interpolate the coarse values of $\varphi$ onto the center of the fine cells. The fifth force on the refinement levels is obtained by finite-differencing using the interpolated values of $\varphi$. Naturally, this scheme introduces an error in the absolute value of the fifth force computed at each time step, but the important point is that if the error is induced on a fifth force that is small, then this should not propagate into a noticeable error in the total force. To be more explicit, we can define the effective gravitational strength with and without truncation, respectively, as
\bq\label{eq:errors}
G_{\rm eff}^{trunc.} &=& 1 + \frac{F_{5{\rm th}}^{trunc.}}{F_{\rm GR}}, \\
G_{\rm eff}^{full} &=& 1 + \frac{F_{5{\rm th}}^{full}}{F_{\rm GR}},
\eq
where $F_{5{\rm th}}^{full}$ denotes the fifth force obtained if one iterates the scalar field on the refinements and $F_{5{\rm th}}^{trunc.}$ denotes the value obtained by truncating the iterations on the refinement and taking the interpolated values from the coarse level. We can define the error $\varepsilon^{trunc.}$ induced on the fifth force at each time step as $F_{5{\rm th}}^{trunc.} = F_{5{\rm th}}^{full}\left( 1 + \varepsilon^{trunc.}\right)$, in which case we have
\bq\label{eq:errors2}
G_{\rm eff}^{trunc.} &=& G_{\rm eff}^{full} + \frac{F_{5{\rm th}}^{full}}{F_{\rm GR}}\varepsilon^{trunc.} \nonumber \\
&\approx& G_{\rm eff}^{full},\ \ \ \ \ \ {\rm in\ high\ density\ regions}
\eq
That is, the error induced by $\varepsilon^{trunc.}$ gets suppressed by the term ${F_{5{\rm th}}^{full}}/{F_{\rm GR}}$, which is smaller the higher the refinement level. Moreover, note that the value of $\varepsilon^{trunc.}$ is not necessarily large. If the truncation of the iterations occurs only on levels where the scalar field is already well screened (i.e. $\varphi$ is smooth), then the interpolation from the coarse levels gives an acceptable result anyway. In the end, the error induced on the total force (second term on the right-hand side of the first line of Eq.~(\ref{eq:errors2})) is a product of two small terms, which is why one can afford performing such truncations.

A consideration that arises in our method concerns the criterion to determine the refinement level above which one should truncate the scalar field iterations. On the one hand, the cell size of the first truncated level has to be small enough to capture the effects of the screening on small scales. On the other hand, if the refinement level where the truncation starts is too high (small grid size), then the desired increase in the performance of the code gets compromised. The result depicted in Fig.~\ref{fig:nfw} suggests that a possible criterion could be to truncate the iterations on refinements whose cell size is smaller than a given fraction of halo sizes. For instance, for the nDGP model we consider in this paper, the fifth force is already less than $\sim2\%$ of normal gravity for $r \lesssim R_{200}/2$. In Sec.~\ref{sec:results}, we shall compare the performance of different truncation criteria.

Before we proceed, it is important to stress that in our method, the calculation of the normal gravity force as well as particle positions and velocities make use of the full AMR structure as determined by the particle distribution and grid refinement criterion. The loss in resolution occurs only for the fifth force, but as we argued above and show below, this still gives acceptable results. We note also that the implementation of the multigrid acceleration is as normal for the AMR levels where the scalar field iterations are not truncated.

\subsection{Vainshtein vs. Chameleon screening models}\label{sec:chameleon}

The speed-up method described above assumes that there is an exact correlation between highly-screened and highly-refined regions in the simulation box. This is true for Vainshtein models, like nDGP, because the screening efficiency is directly determined by the local matter density\footnote{This is in the sense that, for instance, in Eq.~(\ref{eq:eomvarphi_sph2}), the screening is determined by the combination $r_V^3/r^3 \propto M(r)/r^3$, which has units of density.}, which is what determines the AMR structure. However, the same is not necessarily true for other types of screening mechanism, such as the Chameleon type \cite{Khoury:2003aq}.  Here, we shall not delve into the details of Chameleon screening \cite{2004PhRvD..69d4026K, 2013CQGra..30u4004K}, but in short, its successful implementation depends on the ratio $|\varphi_{\rm in} - \varphi_{\rm \infty}|/|\Psi|$, where $\varphi_{\rm in}$, $\varphi_{\rm \infty}$ are the values of the scalar field inside and outside a given source, and $\Psi$ is its gravitational potential. The dependence on $\varphi_{\rm \infty}$ makes the efficiency of the screening dependent on an interplay between the properties of the source and its environment. More specifically, the screening is more efficient in denser environments. Hence, it could be the case that a highly refined region may be unscreened because it is located in a very low density environment, e.g., a small highly-concentrated halo inside a void. 

Despite being more involved, it may still be possible to design speed-up methods for Chameleon models based on a similar idea. In such methods, the criterion for the truncation would have to be position dependent, i.e., at a given AMR level, the iterations are truncated on some cells, but not on others. A possible first step in such a direction could involve designing a method that iterates the scalar field at a given cell on level $l+1$ (one of the red cells in Fig.~\ref{fig:amr}) if the fifth force on its parent cell (the cell at level $l$ that was refined) is still large. Conversely, if the fifth force on the parent cell at level $l$ is already small, then the iterations on its son cells at level $l+1$ (those that arise after the refinement) can be truncated. The numerical convergence properties of iterations that are done only on some cells at a given refinement level would have to be investigated carefully. We leave the generalization to other models for future work. In the remainder of this paper, we focus on examining the accuracy of the method for Vainshtein screening models. This said, the conclusions of this paper should hold also for simulations of other Vainshtein screening theories such as the Cubic \cite{2013JCAP...10..027B} and Quartic \cite{2013JCAP...11..012L} Galileon models.

\section{Results}\label{sec:results}

In this section, we measure the impact of different choices of the refinement level above which to truncate the iterations of Eq.~(\ref{eq:newton}). We start by summarizing the specifications of our simulations and illustrate how the fifth-force-to-normal-gravity ratio depends on the refinement level. Then, we compare the results from different truncation criteria for the matter power spectrum and a number of halo properties.

\subsection{Summary of the simulations}\label{sec:sumsim}

\begin{table}
\caption{Summary of the simulations used in this paper. The CPU hour values of the nDGP models are measured w.r.t.~the time taken by the $\Lambda$CDM simulation. All models were simulated using the same number of computer cores. The values of $h_c$ denote the grid size of the refinement level (in ${\rm Mpc}/h$), $l_{trunc}$, above which the iterations of the scalar field are truncated and $l_d$ is the domain level defined as $N_p^{1/3} = 2^{l_d}$. If $l_{trunc} = l_d + 1$, this means that the scalar field is only iterated on the domain level and interpolated to all refinement levels. For all simulations, the grid refinement criterion is $N_{\rm ref} = 4$.}
\begin{tabular}{@{}lccccccccccc}
\hline\hline
Model &\ \  Truncation &  \ \ CPU hours 
\\
\hline
\\
&  $\ \ \ \ L = 250\ {\rm Mpc}/h,\ N_p = 512^3, l_d = 9$
\\
\\
$\Lambda$CDM &\ \  $---$ &  \ \ $6643$ 
\\
nDGP &\ \  Full (no truncation) &  \ \ $13.5\times \Lambda{\rm CDM}$
\\
nDGP &\ \  $h_c \leq 0.03\ (l_{trunc} = l_d + 4)$ &  \ \ $9\times \Lambda{\rm CDM}$
\\
nDGP &\ \  $h_c \leq 0.12\ (l_{trunc} = l_d + 2)$ &  \ \ $3\times \Lambda{\rm CDM}$
\\
nDGP &\ \  $h_c \leq 0.24\ (l_{trunc} = l_d + 1)$ &  \ \ $1.5\times \Lambda{\rm CDM}$
\\
\hline
\hline
\end{tabular}
\label{table:boxes}
\end{table}

Our simulations were performed with a version of the {\tt ECOSMOG} code \cite{2012JCAP...01..051L, baojiudgp} that has been suitably modified to take into account the speed-up method described above. We use a simulation box with side $L = 250\ {\rm Mpc}/h$ with $N_p = 512^3$ dark matter tracer particles. In {\tt ECOSMOG} (as in {\tt RAMSES}), the domain level of the AMR structure is characterized by $2^{l_d} = N_p^{1/3} \rightarrow l_d = 9$ and its cell size is $L/2^{l_d} \approx 0.48\ {\rm Mpc}/h$. This is the same box size used in the modified gravity code comparison project of Ref.~\cite{codecomp}. In our simulations, the AMR cells are refined whenever the effective number of particles contained in its spatial volume exceeds $N_{\rm ref} = 4$ (in Ref.~\cite{codecomp}, the grid resolution was lower, $N_{\rm ref} = 8$). We interpolate the density and forces back-and-forth between the particles and the AMR cells using a cloud-in-cell (CIC) interpolation scheme. The cosmological parameters adopted are the same as one of the nDGP runs performed for the code comparison project of Ref.~\cite{codecomp}: $r_cH_0 = 1$, $\Omega_{m0} = 0.269$, $\Omega_{\Lambda, 0} = 1 - \Omega_{m0}$, $h = 0.704$, $n_s = 0.966$, $\sigma_8 = 0.8$, where $\Omega_{\Lambda, 0}$ is the present-day fractional energy density of the cosmological constant (cf.~Sec.~\ref{sec:dgp}), $n_s$ is the spectral index of the primordial scalar fluctuations power spectrum and $\sigma_8$ is the variance of the density field on scales of $8\ {\rm Mpc}/h$. This yields a particle mass of $M_p = \bar{\rho}_{m0}L^3/N_p \approx 8.8\times10^{9}\ M_{\odot}/h$ in our simulations.

In the implementation of our speed-up method, we consider three truncation criteria. We perform simulations of the nDGP model in which the iterations are truncated on levels $l \geq 10$, $l \geq 11$ and $l \geq 13$. For these three cases, the cell sizes at the first truncated level are $L/2^{10} \approx 0.24\ {\rm Mpc}/h$, $L/2^{11} \approx 0.12\ {\rm Mpc}/h$ and $L/2^{13} \approx 0.03\ {\rm Mpc}/h$, respectively. In this paper, we label each of these truncated runs by these cell sizes. The vertical lines in the lower panel of Fig.~\ref{fig:nfw} indicate these values, which lie inside the radial scales where the screening starts to operate for the halo masses shown. Note that the truncated run at $h_c = 0.24\ {\rm Mpc}/h$ corresponds to iterating the scalar field only on the domain level, which is the most aggressive truncation one can do. For comparison, we have also run a {\it full} nDGP simulation, in which the scalar field is iteratively solved on every refinement level, and a standard $\Lambda{\rm CDM}$ simulation. All simulations start from the same set of initial conditions at $z = 49$.

Table \ref{table:boxes} summarizes the simulations that we use in this paper. The CPU hours column displays the increase in the performance of the truncated nDGP runs, measured in units of the time taken by the $\Lambda{\rm CDM}$ run. {\it The truncated run $\tfive$ is one order of magnitude faster than the full nDGP simulation, and only 50\% slower than $\Lambda$CDM. This is one of the main results of this paper.} In the next section, we analyse the simulation results to show that the increase in performance does not come at the price of a noticeable loss in accuracy.

\subsection{Fifth force as a function of the refinement level}\label{sec:ratio}

\begin{figure}
	\centering
	\includegraphics[scale=0.44]{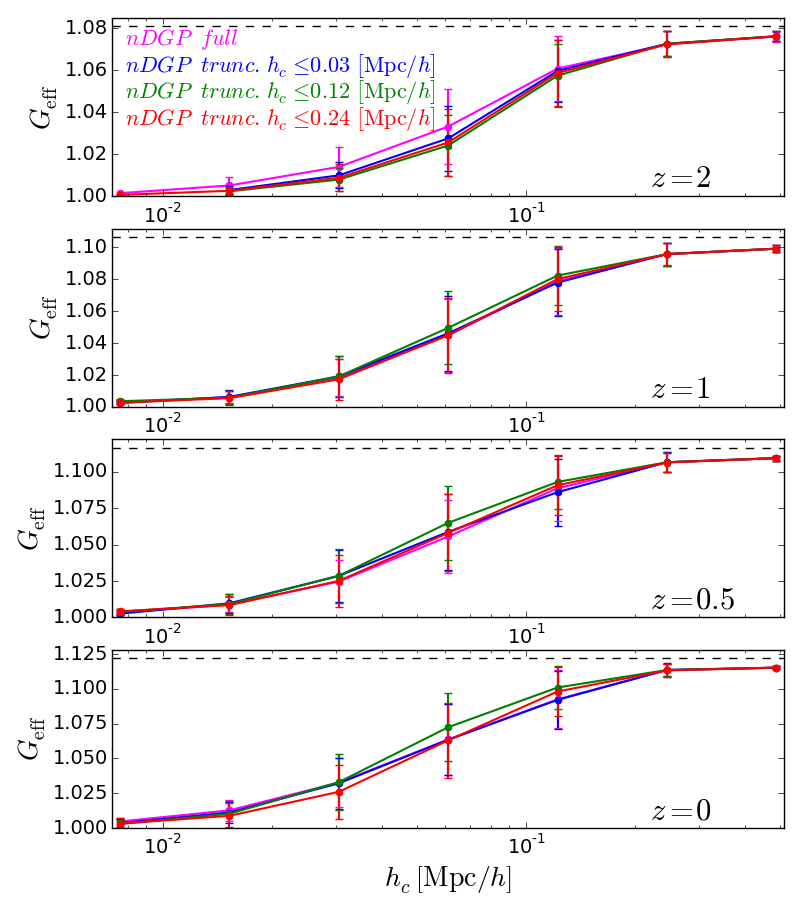}
	\caption{Fifth-force-to-normal-gravity ratio at particle positions as a function of the cell size of the finest refinement level the particles are in, for four epochs ($z = 2,1,0.5, 0$) and for the four nDGP simulations, as labelled. In each panel, from left to right, the points correspond to the AMR levels $l=15,14,13,12,11,10,9$, with the last being the domain level. The points show the mean value of all the particles taken from a slice of $\Delta z_{coord} = 1\ {\rm Mpc}/h$ and the errorbars show the standard deviation around the mean. The horizontal dashed lines indicate the linear theory (unscreened) result.}
\label{fig:ratio}
\end{figure}

Figure \ref{fig:ratio} shows $G_{\rm eff} = 1 + F_{5{\rm th}}/F_{\rm GR}$ evaluated at particle positions in a slice $\Delta z_{coord} = 1\ {\rm Mpc}/h$ of the box, plotted as a function of the finest AMR level the particles are in. The $x$-axis shows the grid size of the AMR levels. The figure shows no trend that the truncation is modifying the mean relation between $G_{\rm eff}$ and the AMR levels\footnote{Note that due to the fact that the gravitational force is not exactly the same in between the different nDGP runs, not all the same particles lie in the same slice of the simulations nor in exactly the same spatial location. Nevertheless, the figure does show that the mean relation is preserved across the different truncated runs.}: all simulations show the expected result that the fifth force becomes weaker in higher refinements, where the density is higher. This agreement holds for all cosmic epochs shown. Note, in particular, that the $\tfive$ truncation agrees very well with the full run, even on the highest refinement levels shown ($h_c \leq 0.01\ \left[{\rm Mpc}/h\right]$). This illustrates that the domain grid captures the effects of the screening sufficiently well {(i.e., $F_{5\rm th}/F_{GR}$ is sufficiently small in Eq.~(\ref{eq:errors2}))} such that there are no marked differences between the result obtained by (i) interpolating the fifth force from the domain to refined levels and (ii) solving the fifth force directly on the refinements. The main difference is that (i) is considerably less computationally demanding than (ii) (cf.~Table \ref{table:boxes}).

\subsection{Matter power spectrum}\label{sec:timeevo}

\begin{figure}
	\centering
	\includegraphics[scale=0.35]{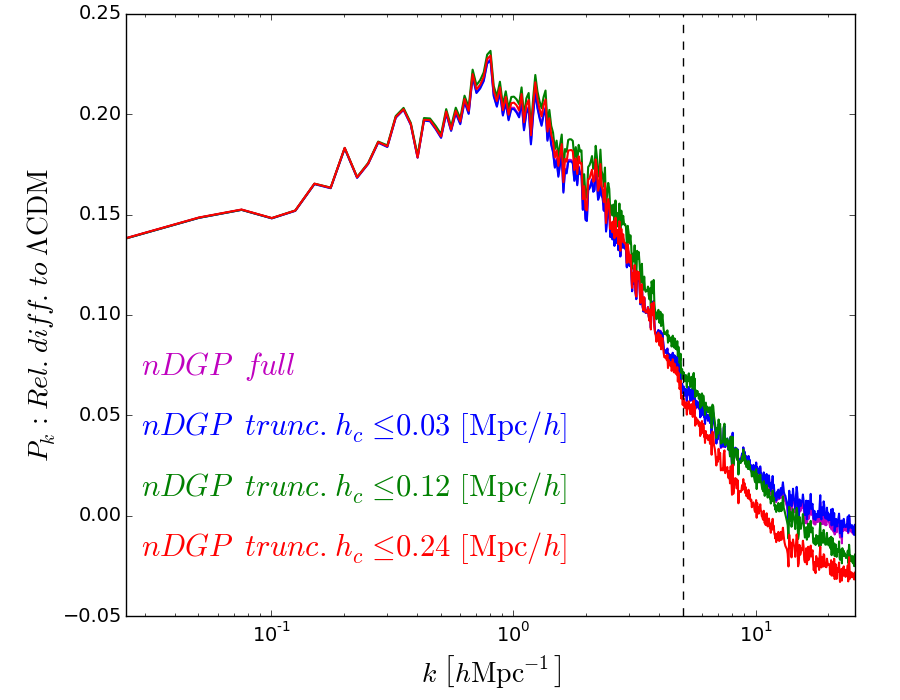}
	\caption{Relative difference of the matter power spectrum, $P_k$, in the nDGP simulations (different colors, as labelled) to $\Lambda{\rm CDM}$ at $z=0$. The vertical dashed line is at $k = 5h/{\rm Mpc}$, which is the $k$-value below which the N-body codes compared in Ref.~\cite{codecomp} agree within $1\%$.}
\label{fig:pk}
\end{figure}

\begin{figure*}
	\centering
	\includegraphics[scale=0.42]{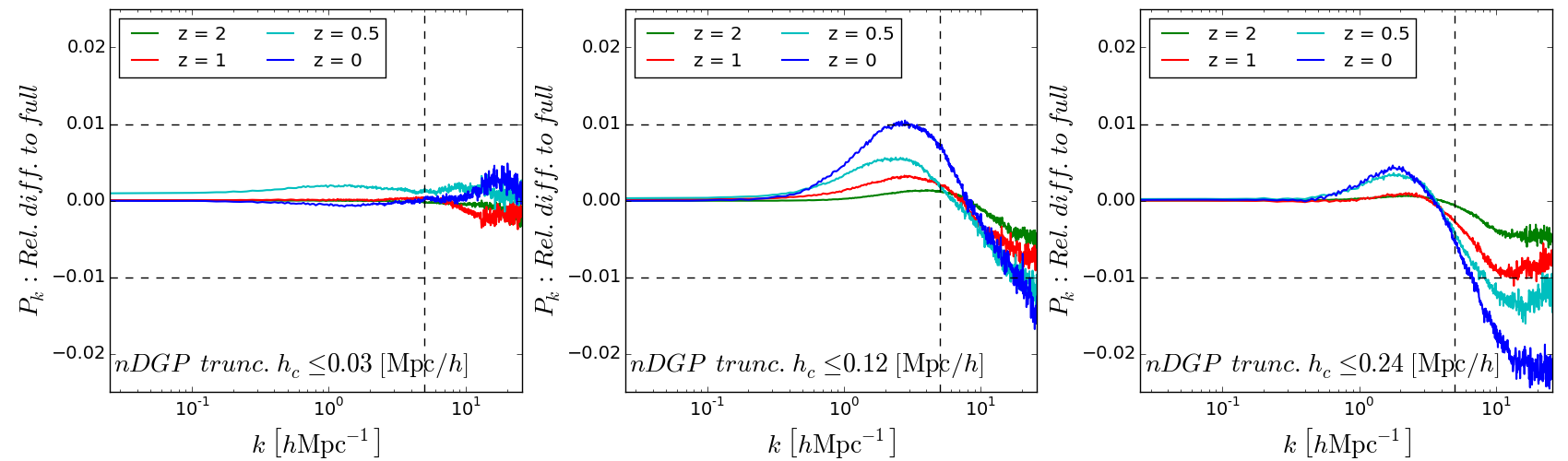}
	\caption{Relative difference of the matter power spectrum, $P_k$, in the truncated nDGP simulations to the full one at $z = 2, 1, 0.5, 0$, as labelled. The three panels correspond to the $\ttwo$ (left), $\tfour$ (middle) and $\tfive$ (right) truncations. The horizontal dashed lines indicate the $1\%$ difference. The vertical dashed line is at $k = 5h/{\rm Mpc}$, which is the $k$-value below which the N-body codes compared in Ref.~\cite{codecomp} agree within $1\%$.}
\label{fig:timeevo}
\end{figure*}

Figure \ref{fig:pk} shows the change in the nonlinear matter power spectrum, $P_k$, in the nDGP simulations w.r.t.~$\lcdm$. We have used the publicly available {\tt POWMES} code \cite{2009MNRAS.393..511C} to compute the power spectrum. The figure displays the known result that the amplitude of the power spectrum is boosted on large scales, $k \lesssim 0.1h/{\rm Mpc}$, and that this boost is amplified on mildly nonlinear scales, $0.1h/{\rm Mpc} \lesssim k \lesssim 1h/{\rm Mpc}$. The screening mechanism starts to manifest itself on scales $k \gtrsim 1h/{\rm Mpc}$, which are typically inside dark matter haloes, $R_{200} \lesssim 1\ {\rm Mpc}/h$ (cf.~Fig.~\ref{fig:nfw}). On scales $k \lesssim 5h/{\rm Mpc}$, all the nDGP runs agree very well with one another. However, for $k \gtrsim 5 h/{\rm Mpc}$, some differences become apparent between the different truncation criteria.

To help understand better these differences, we show in the different panels of Fig.~\ref{fig:timeevo} the relative difference of the power spectrum of each truncated nDGP simulation to the full one.  In each panel, the result is shown for four cosmic epochs $z = 2,1,0.5, 0$. The power spectrum in the least aggressive truncation case, $\ttwo$ (left), agrees to well within the $1\%$ level with the full nDGP run, at all times and scales shown. As one would expect, however, the differences become slightly larger for the criteria $\tfour$ (middle) and $\tfive$ (right). In particular, in the truncated runs, one can identify a scale $k_c$ below which ($k < k_c$) the power tends to be higher in the truncated runs, but above which ($k > k_c$) it drops below that of the full simulation. This effect becomes more pronounced at later times. The scale $k_c$ coincides roughly with the scale of the cell size at which the truncation occurs, $k_c \approx 1/h_c$. {In this paper, we do not pursue a detailed investigation of the origin of these small differences in the power spectrum.}

{For the purpose of our analysis here, the important point from Figs.~\ref{fig:pk} and \ref{fig:timeevo} is that, despite some expected differences induced by the truncation, these are never larger than a few percent ($\lesssim 2.5\%$ at $z = 0$) at all scales shown, and that they become smaller at $z>0$. Moreover, the effects of the truncation are kept below the $1\%$ level for $k < 5h/{\rm Mpc}$, which is the $k$ value above which different modified gravity N-body codes start to exhibit differences larger than $1\%$ as well (see Ref.~\cite{codecomp}). In other words, on those small distance scales, the truncation of the iterations of the scalar field adds only an error that is comparable to the already small differences between different numerical implementations of modified gravity. One should add as well that similar $\%$-level changes in the matter power spectrum are expected from employing different AMR refinement criteria.}

\subsection{Halo counts, concentration, spin and profiles}\label{sec:counts}

In this subsection, we measure the impact of the speed-up method on a number of halo properties, namely their abundance, concentration, spin and profiles. All our halo results are for catalogues built with the publicly available spherical overdensity Amiga's Halo Finder ({\tt AHF}) code \cite{2004MNRAS.351..399G, 2009ApJS..182..608K}. The only exception are the results for the subhalo mass function, which were obtained using catalogues built with the {\tt Rockstar} code \cite{2013ApJ...762..109B}.

\subsubsection{Halo and subhalo cumulative mass function}\label{sec:cmf}

\begin{figure*}
	\centering
	\includegraphics[scale=0.45]{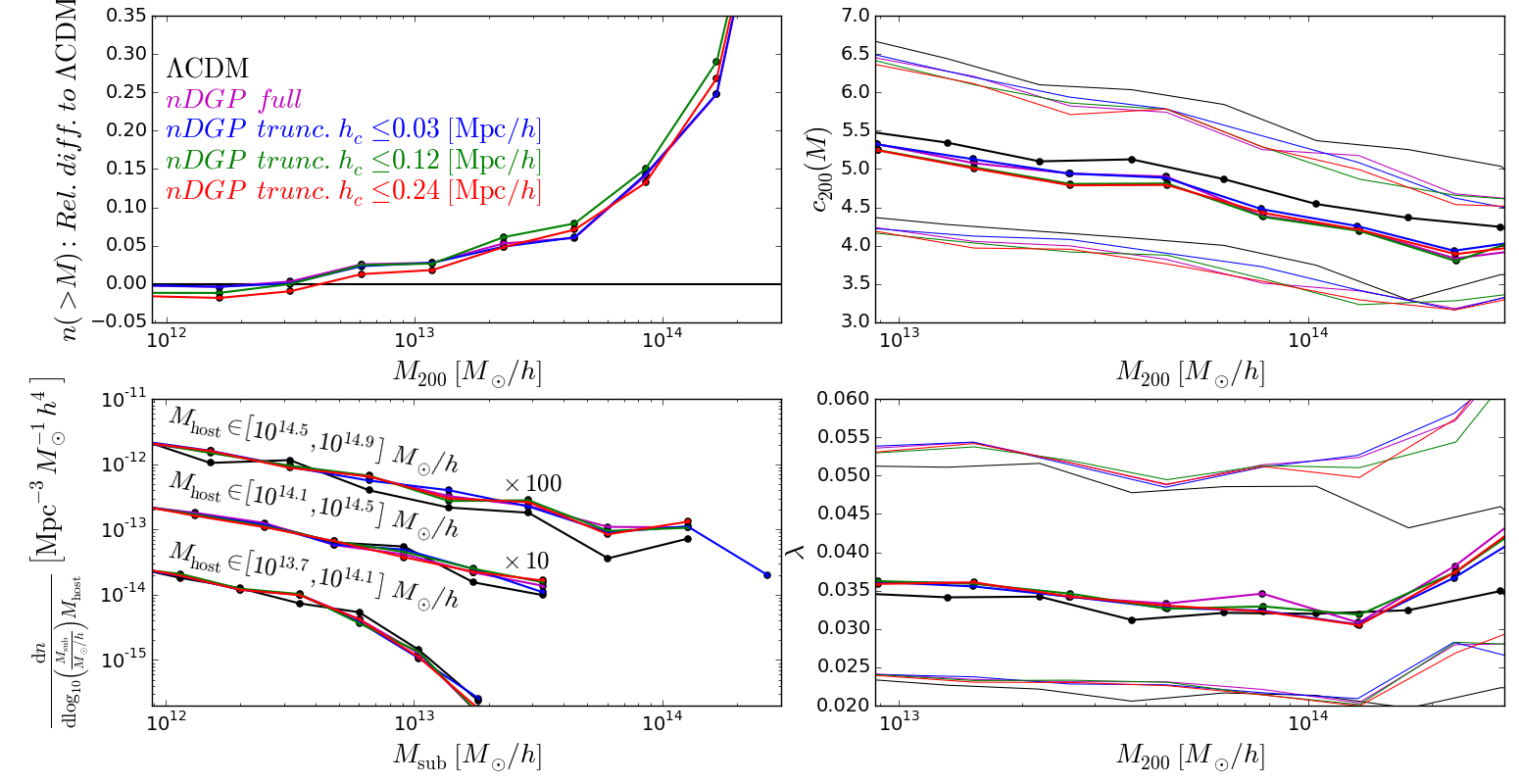}
	\caption{The upper left panel shows the relative difference of the cumulative halo mass function, $n(>M)$, of the nDGP simulations to $\lcdm$. The lower left panel shows the differential subhalo mass function, ${\rm d}n/{\rm dlog}_{10}\left(M_{\rm sub}h/M_{\odot}\right)/M_{\rm host}$, for three host halo mass bins and for the nDGP runs and $\lcdm$, as labelled. The subhalo mass function results of the middle and higher halo mass bins are multiplied by $10$ and $100$, respectively, to avoid overlapping curves. The panels on the right show the mean halo concentration (upper right), $c_{200}(M) = R_{200}/r_s$, and mean halo spin $\lambda$ (lower right), as a function of halo mass (dots connected with solid lines). The thin lines denote the $25\%$ and $75\%$ percentiles of the concentration and spin distributions. We only show the mass function result for haloes with mass $M_{200} > 10^2M_p$, where $M_p = \bar{\rho}_{m0}L^3/N_p \approx 8.8\times10^{9}\ M_{\odot}/h$ is the particle mass. For the concentration and spin, the lowest halo mass shown is $10^3M_p$. All panels are for $z=0$.}
\label{fig:counts}
\end{figure*}

In the upper left panel of Fig.~\ref{fig:counts}, we show the cumulative mass function, $n(>M)$ ($n$ is a number density), of the nDGP simulations, plotted as the relative difference to the $\lcdm$ result. The mass functions show good agreement among the nDGP simulations, with all truncation criteria exhibiting the expected result that massive (low-mass) haloes are more (less) abundant in nDGP gravity, relative to $\lcdm$. The lower left panel of Fig.~\ref{fig:counts} displays the differential subhalo mass function, ${\rm d}n/{\rm dlog}_{10}\left(M_{\rm sub}h/M_{\odot}\right)/M_{\rm host}$, which shows also no evidence that the truncation of the iterations in the nDGP simulations is responsible for making the result less accurate, for all three host halo mass bins shown. Note that there is also no clear difference between the $\lcdm$ and the nDGP prediction for the amount of substructure inside host haloes with $M_{\rm host} \lesssim 10^{14.5}\ M_{\odot}/h$. For $M_{\rm host} \gtrsim 10^{14.5}\ M_{\odot}/h$, the nDGP prediction is slightly but systematically larger than in $\lcdm$ for subhalo masses $M_{\rm sub} \gtrsim 5\times 10^{12}\ M_{\odot}/h$. This could be a result of the effects of the fifth force on large scales, which favours the infall of surrounding low mass haloes into the main more massive one. A more detailed investigation of these effects is beyond the scope of this paper (see e.g.~Ref.~\cite{2015MNRAS.452.3179S} for a more careful study of the subhalo mass function in modified gravity, but in the context of the Hu-Sawicki $f(R)$ model). The important point for our analysis is that any difference to $\lcdm$ is captured consistenly between the different truncated runs.

\subsubsection{Halo concentration and spin}\label{sec:conc}

The right panels of Fig.~\ref{fig:counts} show the mean halo concentration-mass (upper right) and mean halo spin-mass (lower right) relations. Recall that we define the concentration of haloes as $c_{200} = R_{200}/r_s$, where $r_s$ is the best-fitting NFW parameter for a given halo (cf.~Eqs.~(\ref{eq:nfwparams})). The definition of the spin parameter we use is that of Ref.~\cite{Bullock:2000ry}, $\lambda = J_{200}/\left(\sqrt{2}M_{200}V_{c, 200}R_{200}\right)$, where $J_{200}$ is the angular momentum of all halo particles within $R_{200}$ and $V_{c, 200}^2 = GM_{200}/R_{200}$ is the halo circular velocity at $r = R_{200}$. 

On average, when compared to $\lcdm$, the figure shows that haloes in the nDGP model tend to be less concentrated and to spin faster, although the trend may not be very significant when compared, for instance, with the bin-to-bin scatter. When the gravitational strength is boosted, at least two effects arise that can impact on halo concentrations. On the one hand, the stronger gravitational strength makes structures form at earlier times (higher cosmological matter density), and therefore makes haloes more concentrated (see e.g.~Ref.~\cite{2007MNRAS.381.1450N}). On the other hand, particle velocities are also boosted, which can make haloes less concentrated as follows. Consider, for the sake of argument, a collection of particles that are falling radially towards the center of a potential well to form a halo. If the particle velocities get boosted (as they do when gravity is stronger) then the particles should travel a longer distance away from the center after the first {\it shell-crossing}\footnote{We use the term shell-crossing here to denote the stage where the particles cross the center/minimum of the potential well.}. Therefore, at the initial stages of the virialization process, more particles would lie in the outer parts of the halo, making it less concentrated. Hereon, the screening mechanism starts to come into play, and therefore, the modifications to gravity start to impact less the evolution of the halo, which would remain less concentrated. Our concentration results therefore seem to suggest that the effect of boosted velocities wins over the effect of an earlier formation time\footnote{In Refs.~\cite{2011PhRvD..83b4007L, 2011MNRAS.413..262L}, it is also found that the modifications induced by coupled quintessence models can also result in lower halo concentrations.}. This can also help to explain the trend for the spin parameter to be larger in the nDGP simulations compared to $\lcdm$: higher values of the angular momentum could be acquired during the formation of the halo due to the boosted infalling velocities.

A more careful assessement of the mechanisms that determine halo concentration and spin in modified gravity is beyond the scope this paper. For our goals here, what is important is that the different truncated nDGP simulations (including the full one) are consistent in their predictions, which reinforces the validity of our truncation method as a means of increasing the performance of modified gravity N-body simulations.

\subsubsection{Density, velocity and force profiles}\label{sec:profiles}

\begin{figure*}
	\centering
	\includegraphics[scale=0.41]{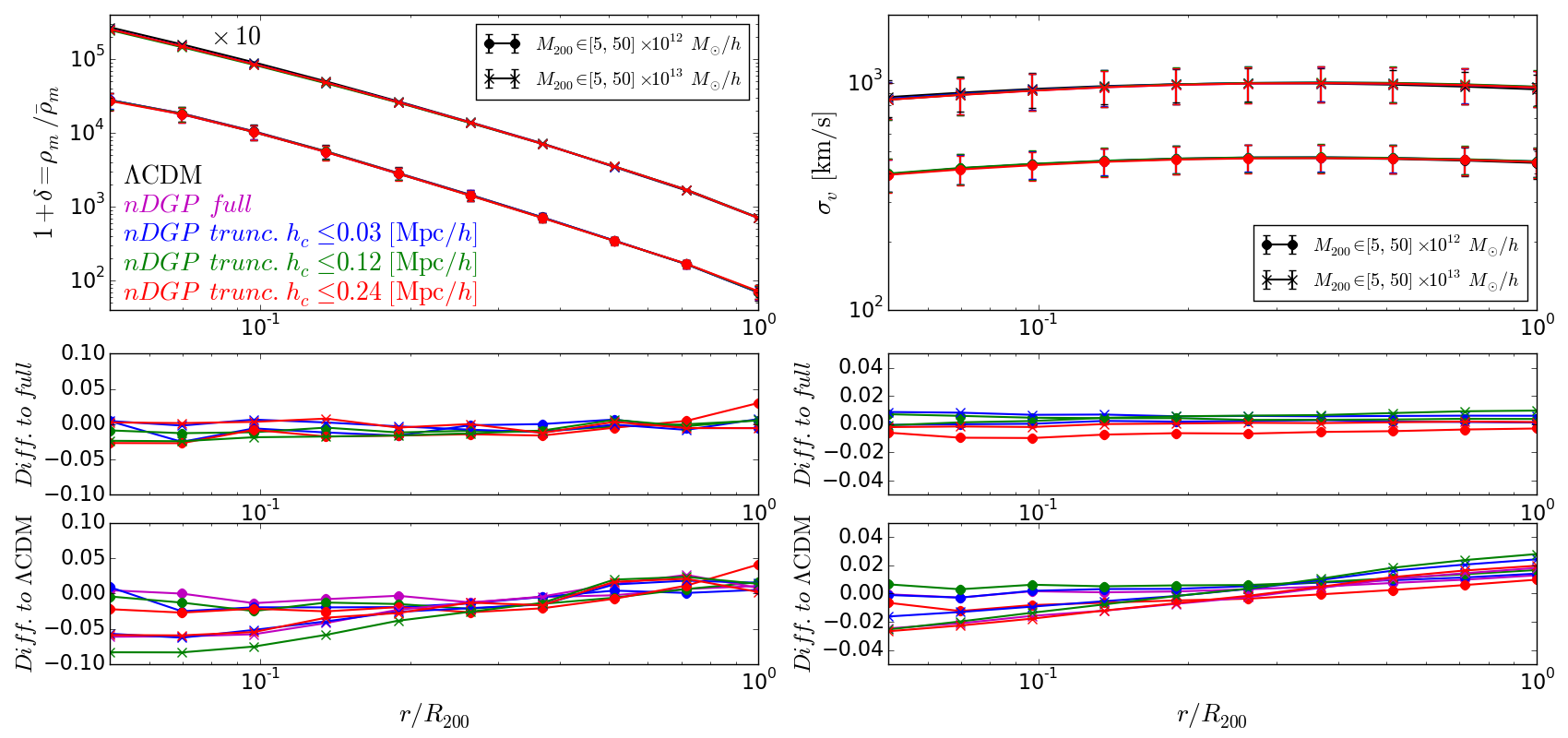}
	\caption{Stacked halo radial profiles for the density, $1+\delta = \rho_m/\bar{\rho}_m$ (left) and velocity dispersion $\sigma_v$ (middle), for two halo mass bins, as labelled. In the upper left panel, the density profiles of the higher mass bin are multiplied by $10$ to avoid overlapping curves with the lower mass bin. The middle panels (from top to bottom) show the relative difference of the truncated nDGP runs to the full one. The lower panels show the relative difference of the nDGP runs to $\lcdm$. The symbols show the mean value for all haloes in the mass bin and the errorbars show the standard deviation. In the panels with relative differences, we do not plot the errorbars for clarity, but we note that in the upper panels, all simulation predictions lie within one anothers errorbars for all radial bins shown. All panels are for $z=0$.}
\label{fig:profiles}
\end{figure*}

The halo radial profiles of the density, $1 + \delta = \rho_m/\bar{\rho}_m$ (left) and velocity dispersion $\sigma_V$ (right) are shown in Figure \ref{fig:profiles} for $z=0$ and for all the nDGP runs and $\lcdm$. The density and velocity profile results from each of the nDGP simulations and $\lcdm$ are within the errorbars of one another, in each mass bin. The second row of panels displays the relative difference of the truncated runs to the full one and shows that the difference is never larger than a few $\%$ at all radii (for clarity, the errorbars are not shown, but these would make the relative difference consistent with zero).

The lower two panels show the relative difference of the density and velocity dispersion profiles of the nDGP runs to $\lcdm$. As mentioned above, any observed difference is not very significant given the size of the errorbars constructed with our halo catalogues. However, if one ignores the errorbars as an exercise, one notes that there is a slight trend for the halos to be less dense in the inner regions, relative to $\lcdm$. This is consistent with the trend that halo concentrations are slightly lower in nDGP gravity, compared to $\lcdm$ (cf.~Fig.\ref{fig:counts}). On the other hand, one may also note a very weak trend for the $\sigma_V$ profiles of the larger mass bin to be lower (higher) in the inner (outer) regions of the halo. We stress, however, that any of these differences is insignificant given the size of the errorbars (see e.g.~Ref.~\cite{2015MNRAS.449.2837G, 2015MNRAS.452.3179S} for studies of halo velocity profiles, but in other modified gravity models).

\subsection{Halo properties of matched haloes}\label{sec:props}

\begin{figure*}
	\centering
	\includegraphics[scale=0.395]{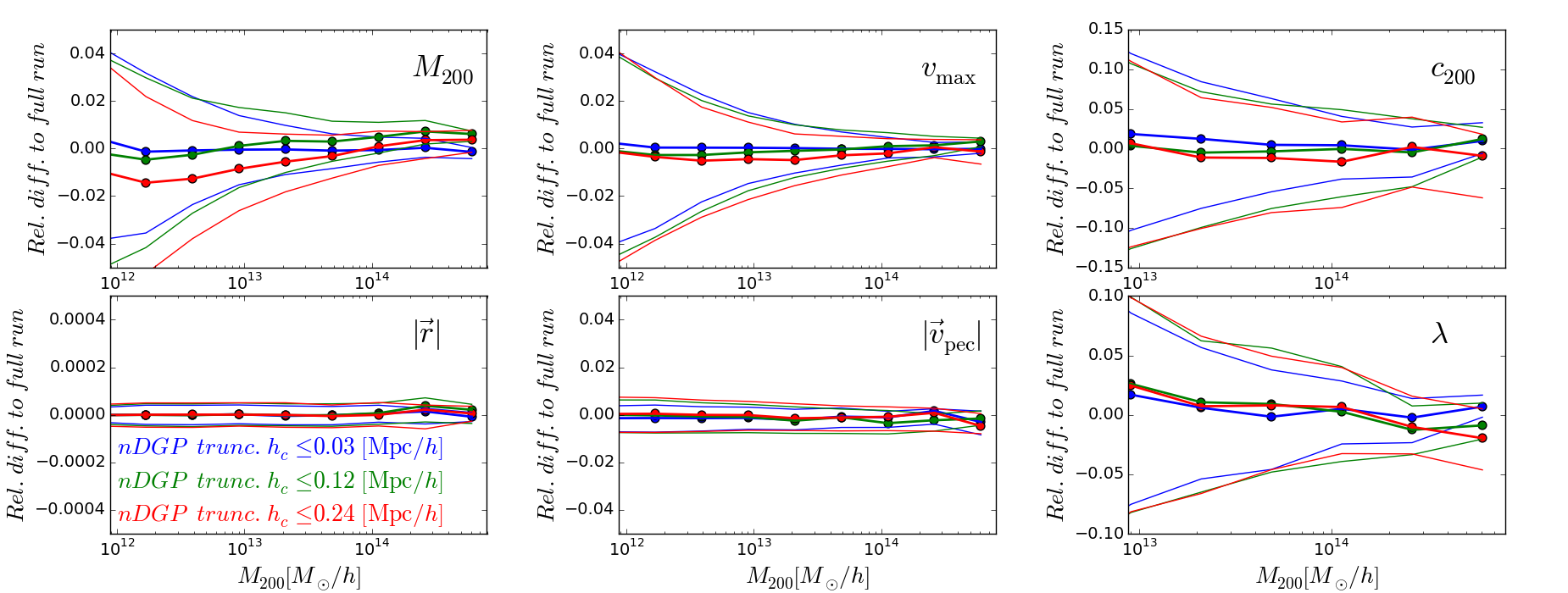}
	\caption{Comparison of halo (not subhalo) properties between the truncated nDGP runs and the full one done on a matched halo basis. The different panels show the comparisons for halo mass, $M_{200}$, maximum circular velocity, $v_{\rm max}$, concentration $c_{200}$, spin $\lambda$, halo position, $|\vec{r}|$ and halo peculiar velocity, $|\vec{v}_{\rm pec}|$, as labelled. For each halo property, the result is the distribution of the relative difference of a given quantity of each pair of matched haloes, plotted as a function of the halo mass in the full nDGP simulation. The dots, linked with the thick solid lines, show the mean value at a given halo mass. The thin solids lines show the $25\%$ and $75\%$ percentiles of the distributions. The minimum halo mass plotted is $10^2M_p$, except in the case of $c_{200}$ and $\lambda$, in which case the mass cut is at $10^3M_p$, where $M_p$ is the particle mass in the simulation.}
\label{fig:props}
\end{figure*}

To complete our assessment of the impact of the truncation of the scalar field iterations in nDGP simulations, we show in Fig.~\ref{fig:props}, a comparison of a number of halo (not subhalo) properties done for matched haloes between each of the truncated nDGP runs and the full one. A halo in one truncated simulation is matched onto the halo that shares the largest fraction of particles in the full run. This one-to-one matching is possible as all simulations start from the same initial conditions. The figure shows the distribution of the relative difference of a given quantity in each pair of matched haloes, plotted as a function of the mass of the halo in the full nDGP simulation.

For $M_{200}$, $v_{\rm max}$, $c_{200}$ and $\lambda$, the mean value of the relative difference never exceeds a few percent ($\lesssim3\%$), on all mass scales shown. In these four panels, the curves for all three truncated nDGP runs are relatively close to each other, except for the case of $M_{200}$, in which the truncation $\tfive$ deviates slightly from the other two at low mass ($\lesssim 10^{13}\ M_{\odot}/h$). We do not find this deviation worrying because it is small ($< 2\%$) and also because a difference of only a few particles in small halos can induce a visible change in mass. For instance, a difference of $20$ ($2$) particles in a $\sim 1000$ ($\sim100$) particle halo can cause an error of $2\%$ in mass\footnote{Note also that the scatter in the distribution at low mass (measured by the $25\%$ and $75\%$ percentiles shown as the thin solid lines in Fig.~\ref{fig:props}) is approximately the same for all truncated runs, including the least aggressive one at $\ttwo$. This hints that the small differences in the properties of matched haloes could be mostly due to the noise that comes from "perturbing" slightly the fifth force calculation, and not due to a systematic error induced by the truncations.}. We expect that these (already) small differences may become even smaller with increased mass resolution (recall that for our simulations the particle mass is $M_p = 8.8\times10^{9}\ M_{\odot}/h$).

The observed differences between the truncated and the full nDGP runs are extremely small for halo center positions ($< 0.01\%$) and for halo peculiar velocities ($< 0.05\%$). This is expected as these halo properties are mostly sensitive to the large-scale behaviour of the fifth force, which is less affected by the truncations. An interesting consequence of this is that quantities such as the halo power spectrum or halo peculiar velocity statistics should be also barely affected by the truncations. This means that, owing to our speed-up method, simulations of nDGP gravity can be used to produce, in a timely manner, large dark matter halo catalogues that can be used in different types of observational studies. These include studies of galaxy clustering via Halo Occupation Distribution \cite{2004ApJ...609...35K} and halo abundance matching techniques \cite{2013arXiv1310.3740K}; or the calibration of methods that are sensitive to halo peculiar velocities such as estimations of the growth rate from Redshift Space Distortions (RSD) \cite{2012MNRAS.420.2102S, 2012MNRAS.426.2719R}, the kinetic Sunyaev-Zel'dovich (kSZ) effect \cite{1972CoASP...4..173S, 2012PhRvL.109d1101H, 2015ApJ...808...47M, 2015arXiv151008844B} or direct measurements of galaxy peculiar velocities \cite{2014PhRvL.112v1102H, 2013AJ....146...86T}. Studies such as these have so far lacked a nonlinear modelling of modified gravity that is as thorough as that available for $\lcdm$, precisely because of the greater computational expense of modified gravity simulations.

\section{Summary and Conclusions}\label{sec:conc}

We have proposed a method to improve the performance of current N-body AMR simulations for modified gravity. In previous works of such simulations, the equation of motion of the scalar field that determines the fifth force is solved on all refinement levels of the AMR structure using iterative algorithms (cf.~Sec.~\ref{sec:itermet}), which makes such simulations very slow compared with $\lcdm$ simulations with the same size and resolution. Our speed-up method is based on truncating these iterations in regions where the screening mechanism suppresses the amplitude of the fifth force to a sufficiently small value. The reason why this does not lead to a degraded accuracy is because the error induced on the fifth force barely propagates into the total force, since it represents only a small modification to gravity in screened regions (cf.~Eq.~\ref{eq:errors2}).

We have applied our method to simulations of nDGP gravity, which employs the Vainshtein screening mechanism to recover GR deep inside dark matter haloes. In this mechanism, the efficiency of the screening in a given region is set by the matter density, which is what directly determines if a given AMR cell gets refined or not. Hence, in Vainshtein screening models, there is a tight correlation between highly screened and highly refined regions and our method can be implemented by simply truncating the iterations of the scalar field on all cells above a pre-specified AMR level. The situation is less straightforward for screening mechanisms with environmental dependence, such as the Chameleon screening. In these, it is not necessarily true that a highly-refined region is screened (cf.~Sec.~\ref{sec:chameleon}), which requires generalizations to the method implemented here. We stress, however, that the principle behind the speed-up method that one can afford less accurate fifth force solutions in screened regions remains valid to all types of screening and that it is just the exact implementation of the method that is model dependent.

To illustrate the validity of the method, we implemented it in the {\tt ECOSMOG} N-body code, and ran test simulations using a box with size $L=250\ {\rm Mpc}/h$, $N_p=512^3$ particles and cell refinement criterion $N_{\rm ref}  = 4$. For this box, the cells on the domain (unrefined) level have a side of $\approx0.48\ {\rm Mpc}/h$. To measure the impact of different choices of the AMR level at which to first truncate the scalar field iterations, we compared a full nDGP run, in which the scalar field is iterated on all AMR levels, with three nDGP runs where the iterations are truncated on refinements whose cell size $h_c$ satisfies $\ttwo$, $\tfour$ and $\tfive$ (cf.~Table \ref{table:boxes}). We have also simulated a $\lcdm$ model and all these simulations started from the same initial conditions. Our main results can be summarized as follows:

\hspace{0.2 cm} $\bullet$ The mean relation between the amplitude of the total force at particle positions and their refinement level is not visibly affected by any of the truncation criteria. In particular, in all nDGP simulations, $F_{5\rm th}$ decreases with the AMR level, $l$, as expected (cf.~Fig.~\ref{fig:ratio}).

\hspace{0.2 cm} $\bullet$ The impact of the speed-up method on the nonlinear matter power spectrum is below the $1\%$ level on scales $k \lesssim 5\ h/{\rm Mpc}$, even for the most aggressive truncation criterion $\tfive$ (cf.~Fig.~\ref{fig:timeevo}). For $20\ h/{\rm Mpc} \gtrsim k \gtrsim 5\ h/{\rm Mpc}$, the differences are larger but smaller than $\approx 2.5\%$, which is comparable to the differences between different N-body modified gravity codes \cite{codecomp}. The impact of the truncations becomes even smaller for $z> 0$ (cf.~Fig.~\ref{fig:timeevo}).

\hspace{0.2 cm} $\bullet$ The different truncated nDGP runs (including the full one) predict essentially the same deviations to $\lcdm$ for halo and subhalo number densities (cf.~Fig.~\ref{fig:counts}), halo concentration- and spin-mass relations (cf.~Fig.~\ref{fig:counts}) and halo profiles of the density and velocity dispersion (cf.~Fig.~\ref{fig:profiles}).

\hspace{0.2 cm} $\bullet$ The mean relative difference of halo properties such as mass, maximum circular velocity, concentration, spin, halo center position and halo peculiar velocity in the truncated nDGP runs to matched haloes in the full one never exceeds $\approx 3\%$. The differences in the positions of halo centers and their peculiar velocities are found to be particularly small, $< 0.01\%$ and $< 0.05\%$, respectively. This means that the truncation of the iterations should have a negligible impact on the halo power spectrum and on RSD or kSZ studies.

\hspace{0.2 cm} $\bullet$ As it is the purpose of the development of this method, the truncation of the scalar field iterations leads to a remarkable improvement in the performance of the code (cf.~Table \ref{table:boxes}). In particular, for the truncation criterion $\tfive$, the nDGP simulation becomes only $\approx 50\%$ slower than $\lcdm$. This represents a boost in speed of approximately a factor of $10$ relative to the full simulation, whilst retaining nearly the same cosmological results. This improved performance relative to full simulations is expected to be even more significant for higher particle resolutions.

\bigskip

Our method allows to push the size and resolution of simulation boxes of modified gravity up to levels that were previously hard to achieve due to limitations of computational resources. For these higher resolution setups, a full simulation will not be available, since that would defeat the purpose of running a truncated run in the first place. Consequently, a natural question that arises is the choice of the first refinement level above which the iterations can be truncated. Our results suggest that a valid implementation of the speed-up method requires only that the cell sizes of the first truncated level are small enough (i.e., the density inside is high enough) for the fifth force to be already sufficiently well screened. A possible way to determine the first truncated level is to compute the profile of $G_{\rm eff}$ in NFW haloes, which can be done analytically, and then find the radius $r_t$, defined as the radius where the fifth force drops below a given fraction of the total force. The truncated levels could then be those whose cell size is $\lesssim r_t$. For instance, the rightmost vertical dotted line in Fig.~\ref{fig:nfw} is located at the value of $r_t$ that would correspond to our most agressive truncation criterion $\tfive$, for which $G_{\rm eff} \lesssim \left\{7\%, 4\%, 2\%\right\}$ for $M_{200} = \left\{10^{12}, 10^{13}, 10^{14}\right\} M_{\odot}/h$, respectively. These figures can serve as a guide for future higher resolution simulations of the nDGP model that employ our speed-up method. 

\begin{acknowledgments}

We thank Tom Theuns for encouraging us to develop this project and for fruitful discussions during its early stages. We also thank Kazuya Koyama and Fabian Schmidt for very useful comments and discussions. We thank Lydia Heck for invaluable numerical support. This work was supported by the Science and Technology Facilities Council [grant number ST/L00075X/1]. This work used the DiRAC Data Centric system at Durham University, operated by the Institute for Computational Cosmology on behalf of the STFC DiRAC HPC Facility (www.dirac.ac.uk). This equipment was funded by BIS National E-infrastructure capital grant ST/K00042X/1, STFC capital grant ST/H008519/1, and STFC DiRAC Operations grant ST/K003267/1 and Durham University. DiRAC is part of the National E-Infrastructure. AB thanks the support from FCT-Portugal through grant SFRH/BD/75791/2011 during the initial stages of this work. SB is supported by STFC through grant ST/K501979/1. BL acknowledges support by the UK STFC Consolidated Grant No. ST/L00075X/1 and No. RF040335.

\end{acknowledgments}

\bibliography{ndgptrunc.bib}

\end{document}